\documentclass[letterpaper,preprint]{web4all_cbt}
\usepackage{hyperref}
\usepackage{xspace}
\usepackage{balance}
\newcommand{\negvspace}{\vspace*{-5mm}}

\newcommand{\vips}{visually impaired persons\xspace}
\begin{document}
%
\toappear{}

\title{Accessibility Evaluation of Computer Based Tests}

\numberofauthors{2}
\author{
\alignauthor
Pawan Kr Patel\\
       \affaddr{Indian Institute of Technology}\\
       \affaddr{Kanpur, India}\\
       \email{patelp@cse.iitk.ac.in}
\alignauthor
Amey Karkare\\
       \affaddr{Indian Institute of Technology}\\
       \affaddr{Kanpur, India}\\
       \email{karkare@cse.iitk.ac.in}
}
\maketitle
\begin{abstract}
Computer-based tests (CBTs) play an important role in the professional career of any person. Universities use CBTs for admissions. Further, many large courses use CBTs for evaluation and grading. Almost all software companies use CBTs to offer jobs. However, many of these CBTs do not take attention to the accessibility barriers for persons with disabilities, specifically \vips. In this paper, we present a study of accessibility barriers in various CBTs as faced by \vips in India. These barriers have been identified by a questionnaire survey approach. Our analysis of the responses shows that most CBTs do not meet the expectations of \vips. We conclude the paper with some recommendations to improve accessibility.
\end{abstract}



\section{Introduction}
In today's world, higher education plays a key role in employment for individuals including persons with disabilities. To get admission in a prestigious institution and procuring a well-paid job, one has to pass through the phase of highly competitive examinations. Due to the increasing number of applicants, the mode of conducting exams is gradually shifting from traditional pen-paper based tests to online computer-based tests (CBTs).  Nowadays, all types of examinations including multiple-choice type, written and programming, are conducted using computers and web-based technologies, referred in this paper as CBTs interfaces.  Utilization of these technologies is quite confined for \vips because of barriers to accessing certain components of the interface. This is due to their inability to perform certain physical activities, for instance, \vips find it difficult to read regular fonts and colored graphs during the examinations. There are plenty of challenges to access the computer-based content, as presented in a study by Hayfa et.\ al.~\cite{webaccess:hayfa}. Hence it is absolutely necessary to remove these accessibility barriers in order to provide \vips with an equal opportunity as that to the persons without any disabilities. 
Many CBT systems do have provisions for providing accommodations to \vips to overcome the accessibility barriers during the exams. In India, there are three commonly used accommodations in CBTs:
\begin{enumerate}
    \item Magnification of the text,
    \item Provision of a scribe and 
    \item Compensatory time to complete the examination.
\end{enumerate}
These methods of accommodations are widely used for many years, and are part of government of India guidelines for conducting written examination for persons with disabilities \footnote{\url{http://www.disabilityaffairs.gov.in/content/page/guidelines.php}}. However, to the best of our knowledge,  systematic study about the effectiveness of these methods has not been conducted yet. 

In this paper, we present a study where we have evaluated the availability of common types of accommodations in CBTs in India and their positive and negative impacts on the performance of \vips.  Based on the study of various accommodations and their limitations, we propose a few recommendations for adopting more technology-based accommodations. These recommendations have arrived from the experiences of \vips who appeared in various CBTs, standard practices and manuals~\cite{imp-tech, xrcvc}. The primary benefits of having accommodations that are technology-based are availability and consistency (see Section~\ref{sec:reco}). 

The paper is organized as follows: In 
Section~\ref{sec:rel-work}, we discuss some work similar to our study. This is followed by the detailed description of methodologies used in the study (Section~\ref{sec:method}). Section~\ref{sec:findings} lists our findings about the challenges faced by \vips, while Section~\ref{sec:reco} presents our recommendations to improve the accessibility of CBTs. We conclude the paper in Section~\ref{sec:concl} and present some directions for future work.

\section{Related Work}\label{sec:rel-work}
Assistive technology plays a key role in enhancing the accessibility of computer based tests as presented in a research study by Hakkinen~\cite{assistive-cbta}. Although technology advances have enabled the access of unlimited educational resources to all including \vips, there are still many accessibility barriers in CBTs. The author describes that  in-order to overcome the accessibility barriers in CBTs, technical accessibility standards have to be created. Hence, to evolve these standards and guidelines related to accessibility of CBTs, empirical studies about accessibility of CBTs need to be conducted. The author has given an overview of accessibility barriers in CBTs, while our empirical study presents the accessibility barriers in CBTs along with the effectiveness of various accommodations provided to \vips.

A typical recruitment process involves many steps from the job search to interviewing. A research study by  Grussenmeyer et. al.~\cite{Grussenmeyer:2017:EAJ:3058555.3058570} presented the accessibility barriers in eleven complementary aspects of a classic recruitment process, starting from the job search, application, exams, interview until on-boarding.  The accessibility challenges in each step of a job recruitment process have been highlighted in this study of eight \vips from the technological background. The exams are a key component for final recruitment in a typical setting. Hence accommodations become critical for the success of persons with disabilities. Authors also highlight the {\em unawareness of examinations authorities} about the availability of accommodations for persons with disabilities during the exam. This paper gives an overview of the accessibility concerns of all the components involved in the recruitment process. On the other hand, our study is a detailed in-depth analysis of a single component, (computer-based) exams, of this research study. 

In educational institutions, persons with disabilities have to perform more administrative tasks as compared to persons without disabilities, such as- to fill accommodation form, to get accommodation letter etc. Accessibility barriers associated with each administrative process are studied by Coughlan and Lister~\cite{Coughlan:2018:AAP:3192714.3192820}.  They analyzed the level of difficulty of every single administrative process and reported the impacts on the individual.  The authors have talked about the barriers in requesting various adjustments in studies, examinations, and day-to-day activities (such as travel and parking),  but, unlike our work,  they do not go into the details of specific barriers in the examinations themselves.

Brajnik and Graca \cite{Brajnik:2018:APH:3192714.3192833} presented the accessibility policies as a fundamental instrument for implementing accessibility solutions. They have surveyed the accessibility policies of twenty universities in Europe and highlighted the need for the development of accessibility policies in higher education institutions to cater to the needs (including exam related needs) of persons (students, staffs, and faculties) with disabilities.  The basis to measure the comprehensiveness and concreteness of policy includes three parameters - role, content, and quality of the policy. In the policy-making scenario, authors reported that web-development, and hence web developers to be one of the important components. 

The study presented in our paper complements  the studies mentioned above \cite{Brajnik:2018:APH:3192714.3192833, Coughlan:2018:AAP:3192714.3192820, Grussenmeyer:2017:EAJ:3058555.3058570, assistive-cbta}. CBTs are one of the most important parts of examinations in educational institutions as well as of recruitment process. Therefore, technical accessibility standards, accessibility policies pertaining to CBTs, accessible administrative process and eventually accessible recruitment process are integral to the professional success of persons with disabilities.

\section{Methodology}\label{sec:method}
We created a Google form comprising of single choice, multiple choice, Likert scale type, and short answer type questions. The questions were designed to access the accessibility barriers in the online competitive examinations and entrance tests. Participants were allowed to fill {\em anonymously} the past experiences of up to three online examinations in a single form\footnote{We allowed data for up to three exams to keep a balance between diversity of exams and candidates. Any participant could have filled multiple forms to give data about more exams, but we can not know about it due to the anonymity of participants.}. 
The complete questionnaire and the raw data collected is available at the project page\footnote{\url{https://www.cse.iitk.ac.in/users/karkare/accSurvey/}}.
\subsection{Survey Design}
Questions in the Google form were divided into three major sections, 
\begin{itemize}
    \item{\em General information about the participants:} Information about the age, disability type, and its extent was included in this section along with the objective of the research study and electronic consent for voluntary participation. To facilitate complete anonymity of individual participants, we have not collected any information (such as. email, phone number, address, profession) about the candidates, which can be used for their identification. The anonymity is to allow the participants to give honest feedback about the availability and expectations of accommodations in the CBTs. We did not ask about the gender since neither the disabilities nor the reasonable accommodations to be provided in the examinations are affected by the gender of the candidates.
    
    \item{\em Examination without any accommodation:} In order to discover the accessibility barriers in the CBT interfaces, participants were asked to share their experiences about accessibility of CBT interfaces without any assistance. The major questions include the name of examination, the extent of readability of questions, reading upper case words and sub-scripts / superscripts, typical components which are inaccessible and reasons behind inability to the questions. This gives an idea of what are the typical barriers faced by \vips in a typical CBT.
    
    \item{\em Examination with accommodations:}
    This part is designed to analyze the effectiveness of the current practices of providing accommodations to persons with disabilities who have the limitation in reading and/or writing. Participants were asked about what all accommodations they would like to use and which of them are available during the exam. Apart from that, we have asked about the effectiveness of very common accommodation such as - scribe, magnification and compensatory time. Moreover, participants were asked about their preferred way of reading and overall rating of CBT interface on a Likert scale of $0$ (Very poor) to $5$ (Excellent).
\end{itemize}
\vspace{0.5cm}
\subsection{Data Collected}

In this study, we have confined ourselves to competitive CBTs in India. Although we have not gathered any geographical information about the participants, still there is sufficient diversity in the participants due to the fact that we had targeted the participants from national forums that include educational institutes like IIT Kanpur\footnote{\url{http://www.iitk.ac.in}}, IIT Delhi\footnote{\url{http://www.iitd.ac.in}}, and registered participants at events like Empower 2018\footnote{\url{http://assistech.iitd.ac.in/empower2018/}} and I-STEM Hackathon 2018\footnote{\url{http://inclusivestem.org/hackathon-2018.html}}. These places/events have the participation of persons with disabilities from all over India. Invitations were sent to registered email ids of persons with disabilities on these forums. We had received responses from $24$ participants prior to $Jan \ 15 \ 2019$, out of which one response has been ignored due to no valid entry in the field 'name of examination s/he had appeared in' . In $23$ responses, six participants have shared their experiences in two different CBTs, one participant filled the survey for three different CBTs and remaining 16 participants had filled the questionnaire for only one CBT.

\subsection{Data Prepossessing}
The very initial step was to clean the data. We followed a semi-automated approach where, we wrote scripts in Python$2.7$  along with a little manual intervention. Here are the steps that were followed for preprocessing:
\begin{itemize}
    \item We have considered each exam as a separate entry (row) for our study purpose. The participants who have shared experiences for more than one CBT, information about such participants was duplicated for each CBT. This simplified our  analysis of the data, while maintaining the completeness and correctness.
    \item Rows having the invalid response to 'Name of Online Exam you appeared in' were ignored straight away since these rows defeat our core purpose of the study.
    \item Entries that do not satisfy the conditional dependency of previously asked question were discarded as it does not make sense to keep them in analysis and to distort the actual result. Here is an example of how we discarded such entries. Consider the following questions, where the second question is dependent on the first:
\begin{enumerate}
    \item Did you get the magnification enabled in the exam?
    \item If you have answered ``yes'' to the previous question. Please let us know to what extent are you able to read the questions/instructions during the exam after enabling the magnification?
\end{enumerate}    
    For the responses that contain a 'no' or 'not applicable' to the first question, any response in the second question was discarded and considered to be not applicable.
\end{itemize}

\begin{table}[t!]
\caption{Number of participants appeared in various online exams}
\label{table:exams-data}
\begin{center}
\begin{tabular}{|c|c|}
\hline
\textbf{Name of Exam}  & \textbf{\# Participants} \\ \hline
GATE                   & 7                             \\ \hline
Bank Exams\footnotemark             & 5                             \\ \hline
JEE-Advanced           & 3                             \\ \hline
JAM                    & 3                             \\ \hline
BITSAT                 & 3                             \\ \hline
Campus Placement Exams\footnotemark     & 3                             \\ \hline
RRB                    & 1                             \\ \hline
SAT                    & 1                             \\ \hline
KVPY                   & 1                             \\ \hline
ICET                   & 1                             \\ \hline
Government recruitment exams                    & 1                             \\ \hline
\end{tabular}
\end{center}
$^8${\scriptsize RRB, SBI PO, IBPS etc.}\\
$^9${\scriptsize Participants had reported various interfaces on which they have appeared for campus placement tests which includes Hackerearth. Hackerrank, codecubes, AMCAT and GUVI.}
\end{table}

\subsection{Data Analysis}
We made use of Python 2.7 modules including pandas, numpy, statsmodels, matplotlib and seaborn to analyze the preprocessed data. Since the participants who responded were all \vips except one with cerebral palsy, we analyzed the dimension of the extent of disability (in this case, vision impairment) and made observations of the type of challenges faced by specific ranges of the extent of disability. We compared accommodations available to the \vips with the expected accommodations they mentioned. This data can be used to find the appropriate modifications in existing practices of accommodating \vips during CBTs.

\section{Findings}\label{sec:findings}
In this section, we presented our observations and findings of this empirical study based on the responses received for our survey until January $15^{th}$, $2019$. Note that while we planned to analyze the extent of disability at five different levels between $0$ and $100\%$, since we did not receive any response in the range of $80\%-90\%$ disability, so we have omitted this category in the further discussion. 

The age distribution of the participants was as follows: mean age was 23.4 years, minimum age was 17 years, maximum age was 35 years, first quantile was 22 years, median was 24  years, and 3rd quartile was 25 years. This reflects that participants had a lot of diversity and they appeared in CBTs either for higher studies or for getting a job.We received data about $11$ different exams from different domains, representative of variations in the interfaces of different CBTs as attempted by \vips in India listed in Table~\ref{table:exams-data}

We analyzed the general accessibility of CBT interfaces, followed by the effectiveness of various accommodations provided to \vips in CBTs and the preferences of various accommodations among \vips.

\subsection{CBT Interfaces}
\begin{figure}[t!]
\centering
\includegraphics[width=\columnwidth]{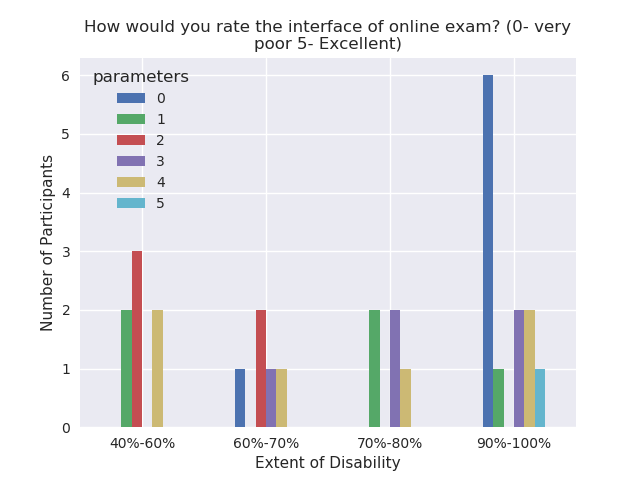}
\negvspace
\caption{Exam interfaces rating}
\label{figure:Exam interfaces rating}
\end{figure}

CBT interface is a combination of many components. Overall accessibility of a CBT interface is determined by the accessibility of individual components. To get a brief idea about the accessibility of typical CBT interface, participants were asked to rate the CBT interface available during CBTs on a Likert scale. The responses can be seen in Figure~\ref{figure:Exam interfaces rating}. A total of $58.6\%$ \vips rated exam interfaces two or below on a scale of $0-5$. In particular $50\%$ of the participants with $90\%$ or more vision impairment gave $0$ score to the CBT interface. This showed the poor experience of \vips with the CBT interfaces. 

\begin{figure}[t!]
\centering
\includegraphics[width=\columnwidth]{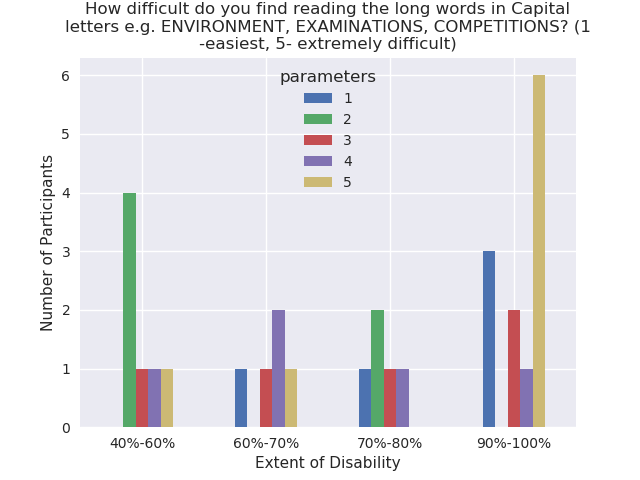}
\negvspace
\caption{Reading the upper-case words}
\label{figure:Reading the upper-case words}
\end{figure}

Figure~\ref{figure:Reading the upper-case words} shows that more than $60\%$ participants faced significant difficulty (rating $3$ or more on scale $1-5$) in reading the upper case words. This observation was consistent with the results in the research done by the cognitive scientists Arditi and Cho.\cite{uppercase:congnitive} on legibility of letter-case for \vips and persons without disabilities.

\begin{figure}[t!]
\centering
\includegraphics[width=\columnwidth]{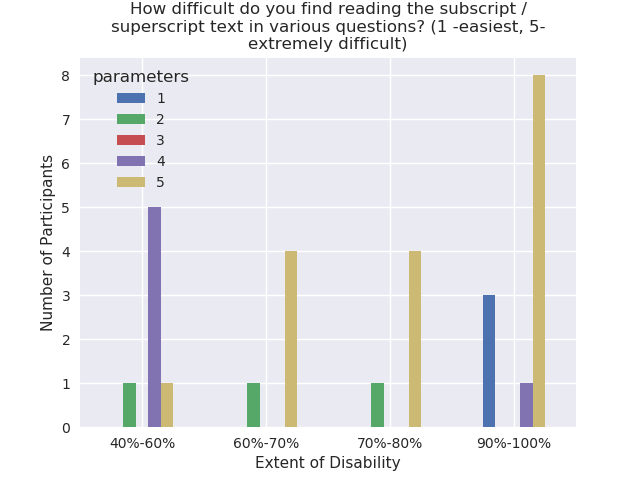}
\negvspace
\caption{Reading the subscripts / superscripts}
\label{figure:Reading the subscripts / superscripts}
\end{figure}

CBTs involving scientific formulae contain subscripts and superscripts. Figure~\ref{figure:Reading the subscripts / superscripts} described the level of difficulty  faced by \vips in reading the subscripts and superscripts. It can be seen that a large majority of the participants reported $4$ or $5$ level of difficulty, irrespective of their extent of disability.

CBT interfaces have other components apart from questions to perform the various operations such as -- selection of the questions, answering the questions, and operating the calculator. We asked the participants to highlight various inaccessible components in a typical CBT interface. Table \ref{table:Inaccessible components of exam interface} listed their responses. It can be seen that more than one-third of participants had some difficulty in accessing the calculator,images and icons embedded in the CBT interface. Two participants reported the barriers in accessing {\em anything that requires movement of the mouse}.

\begin{table}[t!]
\centering
\caption{Inaccessible components of exam interface}
\label{table:Inaccessible components of exam interface}
\begin{tabular}{|c|c|}
\hline
\textbf{Inaccessible Component} & \textbf{\# Participants} \\ \hline
Calculator         & 12                      \\ \hline
Images             & 12                      \\ \hline
Icons              & 10                      \\ \hline
Interaction with Mouse             & 2                       \\ \hline
\end{tabular}
\end{table}

On digging deeper into the reasons behind the accessibility of questions and other components of CBT interfaces, many participants reported some common reasons for inaccessible platforms. These are shown in Table \ref{table:Common reasons for inaccessible interfaces}. We observed that inappropriate font-size used by the CBTs is the basic cause of inaccessibility for approximately $50\%$ of the visually impaired participants. Since \vips have their own font-size reading habits. Providing options for changing the font-size of specific items will help in resolving the major cause of inaccessibility~\cite{uppercase:congnitive} . 

One participant reported a major issue that providing a scribe for a CBT causes him/her to lose any advantage of having a computer-based exam. We quote:
\begin{quote}
GATE $2018$ did not have screen reader installed on their system. And they did not allow it to give it on my laptop. As a result, they provided me scribes. So It apparently makes no difference for me whether it was online or offline. Although the exam was Computer based.    
\end{quote}
This participant raised an important question about the usability of a scribe in the CBT. Whether the exam is paper-based or computer-based, it does not make a difference to him/her because in either case, the scribe reads the content. This reflects, the kind of accommodations provided for paper-based exams are not sufficient for the CBTs.

To discuss the issues with accommodations further, we next analyze the most common ways of accommodations in the exams which includes magnification, scribe and compensatory time. \vspace{0.5cm}

\begin{table}[t!]
\centering
\caption{Common reasons for inaccessible interfaces}
\label{table:Common reasons for inaccessible interfaces}
\begin{tabular}{|l|c|}
\hline
\textbf{Reasons for inaccessibility} & \multicolumn{1}{l|}{\textbf{\# Participants}} \\ \hline
Font size was not appropriate               & 15                                            \\ \hline
Color contrast was poor                     & 9                                             \\ \hline
Lack of screen reader support               & 5                                             \\ \hline
Poor quality of images                      & 1                                             \\ \hline
Surrounding lighting conditions            & 1                                             \\ \hline
\end{tabular}
\end{table}
\subsection{Accommodation: Magnification}
\begin{figure}[t!]
\centering
\includegraphics[width=\columnwidth]{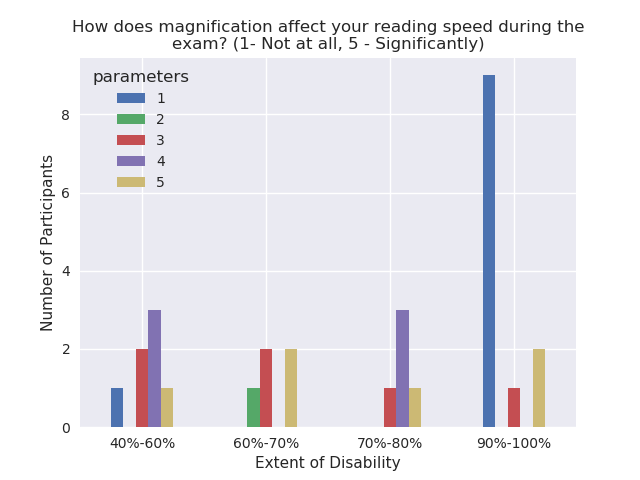}
\negvspace
\caption{Increment in reading speed on magnification}
\label{figure:Increment in reading speed on magnification}
\end{figure}

To evaluate the effectiveness of magnification,  participants were asked to rate the benefit of magnification on Likert scale. The responses are shown in Figure~\ref{figure:Increment in reading speed on magnification}. The effect of magnification depends on the extent of disability, as can be seen from the bar chart. nine out of $12$ people with $90\%-100\%$ visual impairment reported that magnification of the text was barely useful because of less or no residual vision. On the other hand, $15$ out $17$  people with up to $80\%$ disability reported reasonable improvement in reading speed due to magnification. 

\begin{figure}[t!]
\centering
\includegraphics[width=\columnwidth]{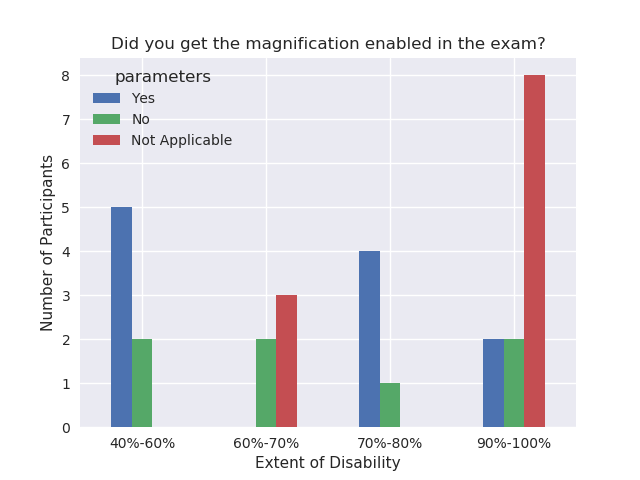}
\negvspace
\caption{Availability of Magnification}
\label{figure:Availability of Magnification}
\end{figure}

On querying about availability of the magnification feature in the examination as showed in Figure~\ref{figure:Availability of Magnification}, only  $37.9\%$ candidates reported that they got magnification enabled in the exam. $24\%$ candidates did not get this facility during the exam, despite their need. Two major reasons behind this could be 1) CBT Authorities did not managed to provide the magnification feature. 2) Participants had not asked for the same. It should also be noted that the candidates who did not ask for this accommodation were mostly in the category of $90\%-100\%$ vision impairment, as explained earlier, magnification accommodation is little or no use for them.

\begin{figure}[t!]
\centering
\includegraphics[width=\columnwidth]{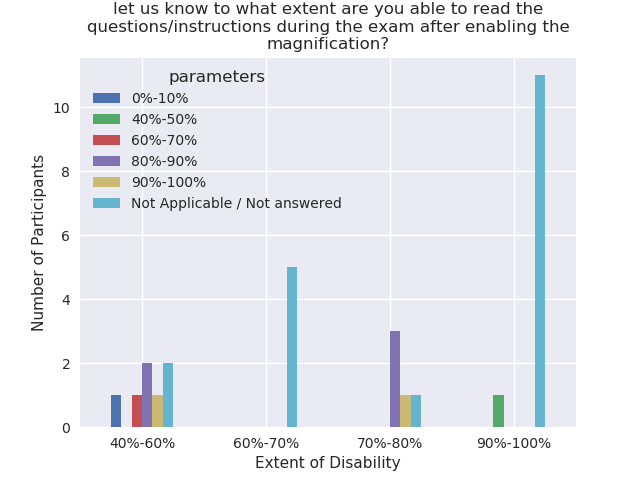}
\negvspace\caption{Extent of reading after enabling magnification}
\label{figure:Extent of reading after enabling magnification}
\end{figure}

\begin{figure}[t!]
\centering
\includegraphics[width=\columnwidth]{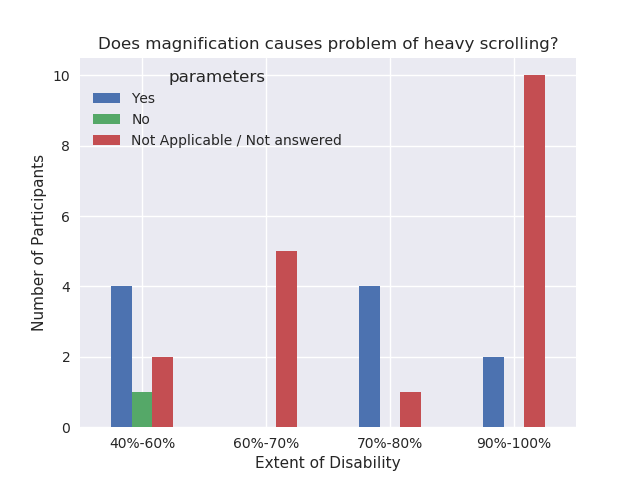}
\negvspace\caption{Heavy scrolling due to magnification}
\label{figure:Heavy scrolling due to magnification}
\end{figure}

\begin{figure}[t!] 
\centering
\includegraphics[width=\columnwidth]{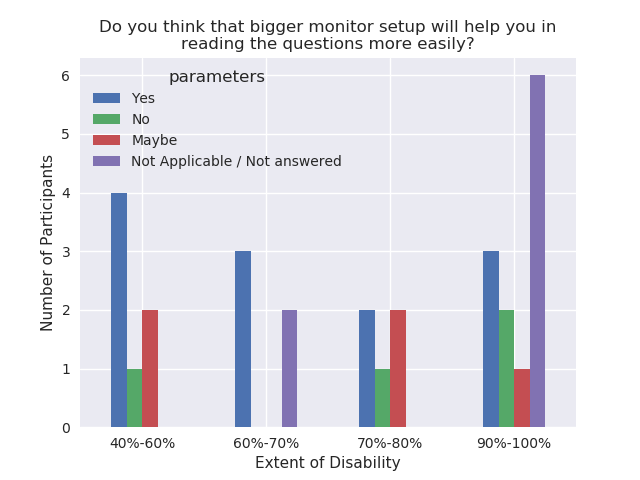}
\negvspace\caption{Possible effect of bigger monitor setup}
\label{figure:Possible effect of bigger monitor setup}
\end{figure}

To evaluate the sufficiency of magnification, participants were asked about the extent of readability of questions after enabling the magnification. Their responses can be seen in the  Figure~\ref{figure:Extent of reading after enabling magnification}. $81\%$ of the people who got the magnification enabled in the CBT were able to read up to $90\%$ or less. This is not sufficient in today's competitive world, where losing a single mark may lead to a lesser grade in a course or disqualification in an entrance. Various reasons due to which participants are unable to read the complete CBT even after enabling the magnification include insufficiency of the extent of magnification and heavy scrolling of text. The problem of heavy scrolling is identified by $90\%$ of the \vips ,who got magnification enabled in CBT can be seen in Figure~\ref{figure:Heavy scrolling due to magnification}.
To enhance the effectiveness of magnification for \vips, the bigger monitor can be provided. This may reduce the need for the large extent of magnification and hence the problem of heavy-scrolling. Responses are given in the Figure~\ref{figure:Possible effect of bigger monitor setup}. More than  $40\%$ \vips opined that bigger monitor setup can help them in easy reading of text. In particular $75\%$ of \vips with $90\%-100\%$ impairment reported that bigger monitor setup is no use for them as enabling magnification does not affect their reading performance.

As we have seen that, accommodation type '\textit{magnification}' was not useful for people with severe vision limitation i.e. more than $90\%$. Hence magnification was not reasonable accommodation for such \vips.As responses from the \vips in the figures from \ref{figure:Increment in reading speed on magnification} to \ref{figure:Possible effect of bigger monitor setup}, shows the majority of them opted for no effect or not applicable.

\subsection{Accommodation: Scribe}
\begin{figure}[t!] 
\centering
\includegraphics[width=\columnwidth]{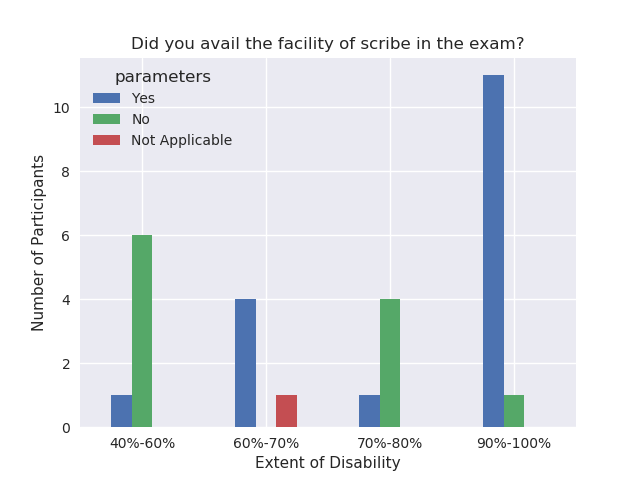}
\negvspace\caption{Availability of Scribe}
\label{figure:Availability of Scribe}
\end{figure}

A Scribe is another most commonly used accommodation in CBTs to provide equal opportunity to \vips. Figure~\ref{figure:Availability of Scribe} displays the distribution of scribe accommodation used by \vips. The participants who have $90\%$ or more disability out of these, $91.6\%$ used a scribe in the CBTs. On the other hand participants having $80\%$ or less vision impairment, only $35.2\%$ used a scribe in the CBTs. Participants with $90\%$ or more disability preferred to use scribe due to lack of residual vision in their eyes, hence they require assistance in reading questions/instructions during the CBTs. While participants with residual sensitivity in their eyes can read the information with sufficient magnification and/or appropriate colour contrast.

\begin{figure}[t!] 
\centering
\includegraphics[width=\columnwidth]{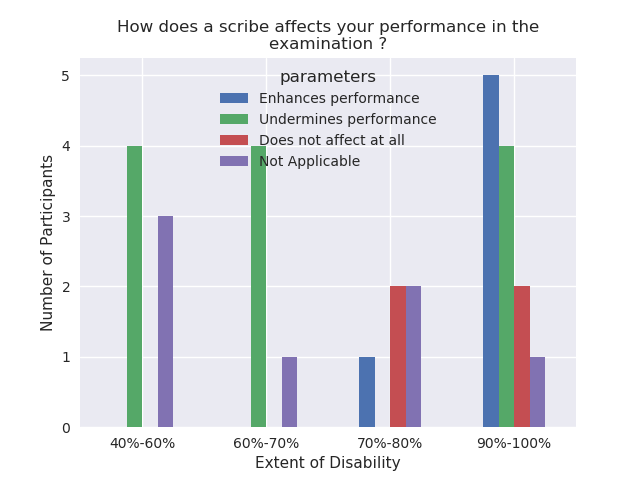}
\negvspace\caption{Impact on performance by utilizing scribes}
\label{figure:Impact on performance by utilizing scribes}
\end{figure}

In order to analyze the effectiveness of the scribe accommodation in CBTs, participants were asked to report the impact of scribe assistance on their performance in the CBT. It can be deduced from the  Figure~\ref{figure:Impact on performance by utilizing scribes} that only $27.3\%$ participants out of those who used a scribe during the CBTs reported enhancement in their performance. Rest of the participants reported either drop or no impact in their performance in CBTs due to scribes. Let consider the participants having $90\%$ or more disability, Now from the Figure~\ref{figure:Impact on performance by utilizing scribes} it is easy to see that $41.6\%$ of such participants who used a scribe reported enhancement in their performance, while $50\%$ reported drop or no impact on their performance. The participants having  $80\%$ or less disability, $58.8\%$ of them reported drop or no-impact in their performance in CBTs and only $5.8\%$ reported enhancement in the CBT performance. The above statistics of effectiveness a scribe in CBT showed that accommodation scribe is beneficial for limited set of \vips who have very less or no residual sensitivity in their eyes.

The discussion above brought us to a conclusion that \vips who can read with the magnification do not prefer to use scribe for assistance in reading. To validate our reasoning, we asked the participants to provide feedback about their experiences with the scribe during CBT. The major concerns of the participants were as follows:
\begin{itemize}
    \item Difference in reading style of scribe and the participants expect to receive the information.
    \item Inadequate competence level of the scribe as per the examination requirement, to read the questions and notations with appropriate context.
    \item Language and accent barriers.
    \item Lack of practice with a scribe in the day to day life.
    \item Lack of independence and control over the exam.
\end{itemize}
We quote some of the comments provided by the participants who used a scribe during CBT:
\begin{quote}
    Screen reader provides independence to re-read and navigate quickly as compare to scribe, especially for the long question, I am habitual of using screen reader as compare to scribe as I rarely get a chance to practice with a scribe.
\end{quote}
\begin{quote}
    It's some time feel better to do to the accessibility of equations and other symbols, but assistive technology is better if there is good access.
\end{quote}
\begin{quote}
    The scribe was less well versed with English and often did mistakes. He was also unaware of many maths symbols. Also as he was always looking into my paper I was losing confidence in myself.
\end{quote}
\begin{quote}
    In normal day to day life, I am habituated to reading by myself rather than by a scribe. Hence it will have a bearing on the swiftness of understanding the question.
\end{quote}
\begin{quote}
    Its hard to understand the code being read aloud to me.
\end{quote}

\subsection{Accommodation: Compensatory time}
\begin{figure}[t!] 
\centering
\includegraphics[width=\columnwidth]{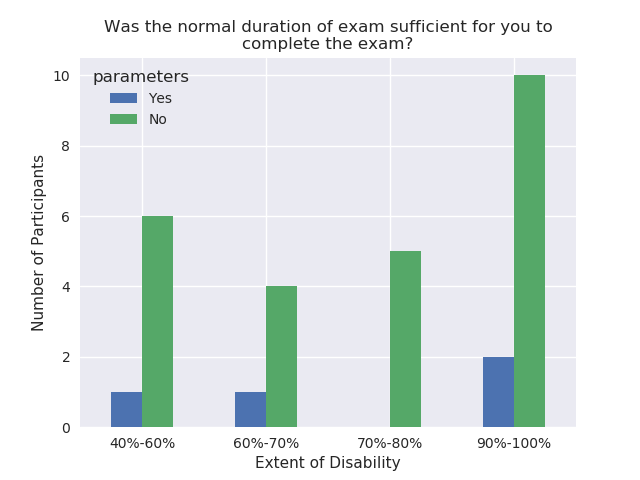}
\negvspace\caption{Sufficiency of standard time duration}
\label{figure:Sufficiency of standard time duration}
\end{figure}

\begin{figure}[t!] 
\centering
\includegraphics[width=\columnwidth]{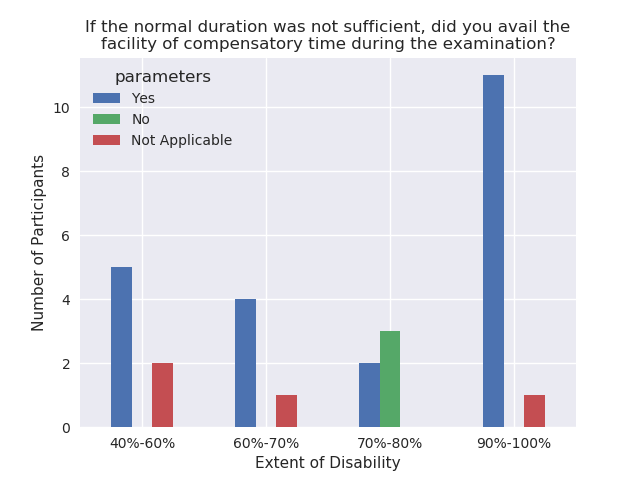}
\negvspace\caption{Availability of compensatory time}
\label{figure:Availability of compensatory time}
\end{figure}
Compensatory time is very important accommodation which compensate the time consumption in the CBTs due to various accessibility barriers by providing extra time to complete the exam. Intuitively, compensatory time would be a need of every visually impaired person because of some difficulty in reading the question/instructions from screen. We evaluated the sufficiency of standard time duration in the exam. As it can be seen in Figure~\ref{figure:Sufficiency of standard time duration} that  $86.2\%$ participants reported standard time duration was not sufficient to complete the exam. This evidence confirmed our intuition about the limitation of the magnification and scribe. In order to get more details, participants were asked  if compensatory time was beneficial for them or not. From Figure~\ref{figure:Availability of compensatory time} it can be observed that regardless of the extent of visual impairment, participants used the compensatory time. In total $75.8\%$ participants reported the same.

\begin{figure}[t!] 
\centering
\includegraphics[width=\columnwidth]{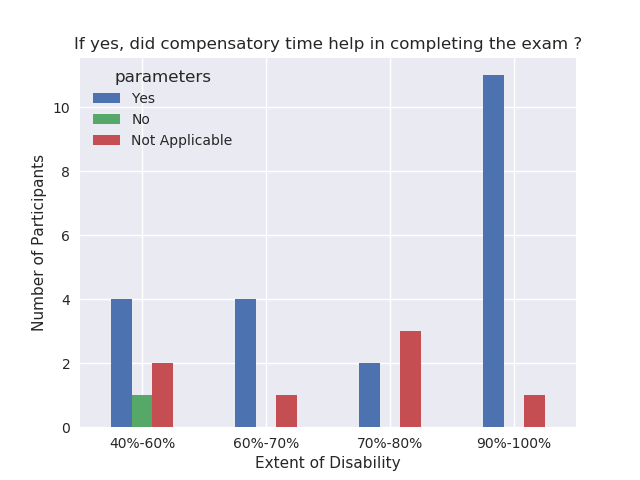}
\negvspace\caption{Efficiency of compensatory time}
\label{figure:Efficiency of compensatory time}
\end{figure}

Figure~\ref{figure:Efficiency of compensatory time} shows that $95.4\%$ of those who had availed the facility of compensatory time reported that it indeed helps in completing the examinations.

\subsection{Preferences among accommodations}
\begin{figure}[t!] 
\centering
\includegraphics[width=\columnwidth]{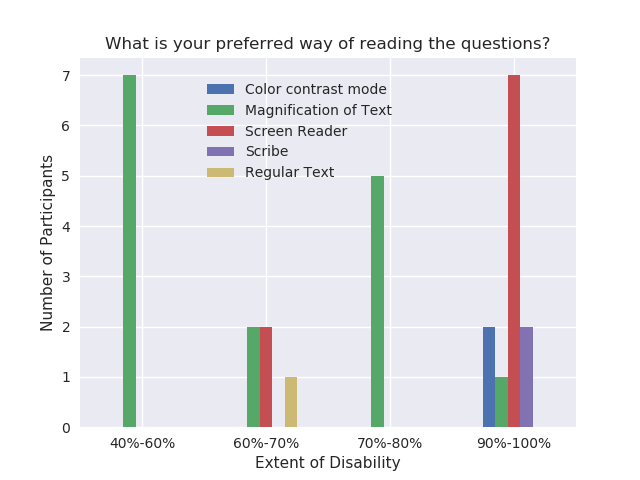}
\negvspace\caption{Reading preferences}
\label{figure:read-pref}
\end{figure}

We saw that different type of accommodations were suitable for the different set of individuals depending on the extent of their disability. To further illustrate the fact, we asked participants about their preferred way of reading the questions. As it can be seen from  Figure~\ref{figure:read-pref}, $82.3\%$ \vips having disability $80\%$ or less, prefer to read text with the help of magnification. Very few of them use a screen reader. On the other hand, $58.3\%$ of \vips having more than $90\%$ vision impairment prefer to read with a screen reader. It is interesting to note that only $6.9\%$ \vips prefer to use scribe over other types of accommodations. 

\begin{figure}[t!] 
\centering
\includegraphics[width=\columnwidth]{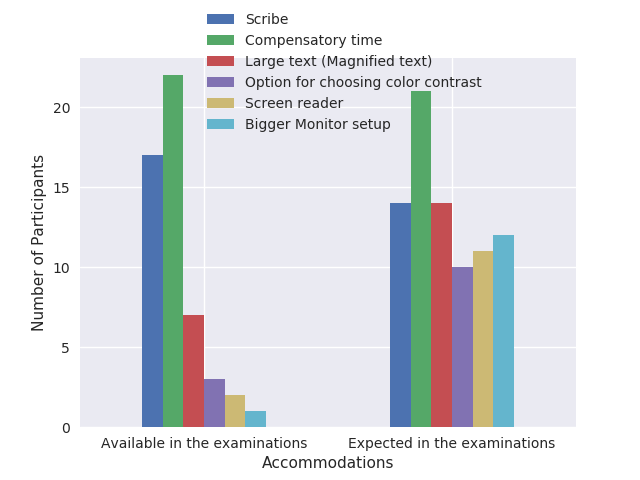}
\negvspace\caption{Available vs expected accommodations}
\label{figure:avail-accmmod}
\end{figure}

In reality, a single type of accommodation does not suffice to overcome the accessibility barriers in the CBTs. Hence we did a comparison between the  accommodations available and accommodations expected by the participants. The responses were showed  in the figure~\ref{figure:avail-accmmod}. Availability and expectation of scribe and compensatory time are very similar and these accommodations used by $50\%$ and $70\%$ of the \vips. On the other hand  accommodations such as: magnification of text, screen reader, option for choosing color contrast and bigger monitor setup were rarely available during CBTs, but $30\%-50\%$ \vips needed these accommodations.

\section{Our Recommendations}\label{sec:reco}
Accessibility of CBTs is crucial for the good performance of \vips. Our survey shows that there are limitations of commonly used accommodations, such as lack of competence level of the scribe, insufficient magnification, inaccessible mathematical content and diagrams for screen reader users. In this section, we propose few recommendations to improve the accessibility of CBTs. These recommendations are based on the feedback provided by the participants of our survey, related work (Section~\ref{sec:rel-work}) and other resources~\cite{imp-tech,xrcvc}.

In general, technology-based assistance has a major role in independent living~\cite{imp-tech} of the \vips. Technology-based assistance is preferred by \vips because of consistency in support and assurance of availability, as compared to human-based assistance (such as scribe) lacks consistency (for example, having a different scribe for different exams). In recent years, with the advancement in the assistive technology,  \vips are able to perform reading and writing tasks on their own, subject to the availability of the contents in an accessible format.

As shown in the research study by Web-Aim\footnote{\url{https://webaim.org/articles/visual/}}, \vips can experience different kinds of accessibility barriers due to variations in the causes of visual impairment. Hence there is a need for adaptive accessible CBT interface, which has the capability to personalize the CBT as per the requirement of individuals.
Since every visually impaired person has a unique set of requirements in terms of accommodations in the CBTs, so CBT conducting authorities should ask the visually impaired persons about their requirement of accommodations at the very beginning of application form filling process. In order to achieve this, there is a need for formation of accessibility policy for CBTs, which entertains the accessibility and universal design of CBT interface.

To improve the effectiveness of existing accommodations for \vips, we propose the following\linebreak accommodation-specific recommendations:
\begin{itemize}
    \item \textbf{Scribe:} To minimize the dependency on scribe, the following technology-based solutions can be used:
    \begin{itemize}
        \item Use of screen readers such as JAWS\footnote{Job Access With Speech: \url{http://www.karishmaenterprises.com/JAWS.htm}}, NVDA\footnote{Non Visual Desktop Access - \url{https://www.nvaccess.org/}}. 
        \item CBT questions in text-based DAISY\footnote{\url{https://nfb.org/images/nfb/publications/bm/bm11/bm1102/bm110210.htm}} format so that screen reader users can access the content.
        \item Providing CBTs in an alternate format such as computer voiced test similar to the one already available for GRE\footnote{\url{https://www.ets.org/research/topics/assessing_people_with_disabilities}}
    \end{itemize}
    \item {\bf Highlighting: } In our survey, $62\%$ \vips reported average or more difficulty in reading the upper-case sentences and words. Upper-case words or sentences are used to highlight important things in the text. Here are some alternatives those can be used in the place of upper-case words:
        \begin{itemize}
            \item Increase the font size of the text
            \item Use of bold font-face
            \item Use of camel case words
            \item Highlighting of text background with different color
            \item Create a box around the text
            \item Underlining of the text
            \item Change of font-family can also be a good option for highlighting.
        \end{itemize}
        To achieve the objective of highlighting the text and its accessibility, the suitable combination of above can be used. Since the combination of above may varies person to person, hence system should have the capability to set the appropriate combination prior to the examination.
    \item \textbf{Magnification: } Accessibility of various icons, subscripts and superscripts in the text can be enhanced by providing two kinds of magnification feature in the interface: one which zooms in the whole screen, and the other which zooms only a specific area on the screen to see very small fonts. Such magnifications are already available on popular operating systems like Microsoft Windows and Apple iOS. 
    \item To avoid heavy scrolling, bigger monitors can be provided to \vips. This will be highly beneficial as requested by $41.3\%$ participants in our survey. 
    \item Inaccessibility of calculators can be removed by providing talking calculator\cite{xrcvc}.
\end{itemize}

\section{Conclusions and Future Work}\label{sec:concl}
Every visually impaired candidate is unique and requires appropriate accommodations in CBTs to overcome the accessibility barriers. In this work we have identified the needs for accommodation of a diverse group of \vips. We showed that extent of disability is a major factor in determining the type of accommodation suited for an individual.  With the help of assistive technologies, CBT interface can be personalized as per the need for visually impaired persons.

A possible threat to the validity of our study and recommendations  is the small number of participants. To offset this, we have proposed our recommendations based not only on the feedback by the participants, but also on the standard practices and manuals~\cite{imp-tech, xrcvc}. 

Our current study is focused on accommodations for \vips in India, and the data received is for CBTs having multiple-choice questions. In future, we would like to extend our study for other disabilities, different types of CBTs and for other geographic locations. Moreover, similar studies can be conducted for the accessibility of online programming competitions.
\\

\section{Acknowledgements}\label{sec:ack}
We would to thank all the participants who have spent their valuable time for voluntary participation in our research study.
\balance
\bibliographystyle{abbrv}
\bibliography{web4all_cbt} 

\end{document}